\documentclass{elsart}
\usepackage{graphicx}
\usepackage{natbib}
\usepackage{amsbsy}

\bibpunct{[}{]}{;}{a}{}{,} 

\newcommand{\be}{\begin{equation}}
\newcommand{\ee}{\end{equation}}
\newcommand{\bvt}{\begin{verbatim}}
\newcommand{\evt}{\end{verbatim}}
\newcommand{\bea}{\begin{eqnarray}}
\newcommand{\eea}{\end{eqnarray}}
\newcommand{\bc}{\begin{center}}
\newcommand{\ec}{\end{center}}
\newcommand{\fly}{\texttt{\large FLY\hspace*{2mm}}}
\begin{document}
\runauthor{Antonuccio, Becciani and Ferro}
\begin{frontmatter}
\title{FLY. A parallel tree N-body code for cosmological simulations. Reference Guide}
\author[OACT]{V. Antonuccio-Delogu},
\author[OACT]{U. Becciani} and
\author[UCT]{D. Ferro}

\address[OACT]{Osservatorio Astrofisico di Catania,
              Via Santa Sofia 78, I-95123 Catania, ITALY\\
              e-mail: (van,ube)@sunct.ct.astro.it
             }
\address[UCT]{Dipartimento di Fisica e Astronomia,
              Via Santa Sofia 78, I-95123 Catania, ITALY\\
              e-mail: dfe@sunct.ct.astro.it
             }

\begin{abstract}
FLY is a parallel treecode which makes heavy use of the one-sided communication
paradigm to handle the management of the tree structure. In its public version the code
implements the equations for cosmological evolution, and can be run for different
cosmological models.\\
\noindent
This reference guide describes the actual implementation of the algorithms of the public
version of FLY, and suggests how to modify them to implement other types of equations (for
instance the Newtonian ones).
\end{abstract}

\end{frontmatter}

{\bf PROGRAM SUMMARY}

{\it Title of Program:} \fly

{\it Catalogue Identifier:} 

{\it Distribution Format:}

{\it Computer for which the program is designed and others on which it has
been tested:} Cray T3E, Sgi Origin 3000, IBM SP

{\it Operating systems or monitors under which the program has been tested:} 
Unicos 2.0.5.40 , Irix 6.5.14, Aix 4.3.3

{\it Programming language used:} Fortran 90, C

{\it Memory required to execute with typical data:} about 100 Mwords with 2 
million-particles

{\it Number of bits in a word:} 32

{\it Number of processors used:} parallel program. The user can select the number of processors 
$\ge 1$

{\it Has the code been vectorised or parallelized?: } parallelized

{\it Number of bytes in distributed program, including test data, etc: } 110 Mbytes

{\it Distribution format:} tar gzip file


{\it Nature of physical problem:} \fly is a parallel collisionless N-body code 
for the calculation of the gravitational force

{\it Method of solution:} It is based on the hierarchical oct-tree
domain decomposition introduced by Barnes and Hut (1986)

{\it Restrictions on the complexity of the program:}  The program uses the leapfrog integrator 
schema, but could be changed by the user

{\it Typical running time : } 50 seconds for each time-step, running a 2-million-particles simulation on
 an Sgi Origin 3800 system with 8 processors having 512 Mbytes Ram for each processor

{\it Unusual features of the program :}  \fly uses the one-side communications libraries:
the shmem library on the Cray T3E system and Sgi Origin system, and the lapi library 
on IBM SP system

{\it References :}  U. Becciani, V. Antonuccio {\em Comp. Phys. Com.} {\bf 136} (2001) 54

{\bf LONG WRITE-UP}

\section{Introduction}
\fly is a parallel collisionless N-body code which relies on the hierarchical oct-tree
domain decomposition introduced  in \cite{1986Natur.324..446B} for the calculation
of the gravitational force. Although there exist different publicly available parallel
treecodes, \fly differs from them because it heavily relies on two parallel programming
concepts: {\it shared memory} and {\it one-sided communications}. Both of these
concepts are implemented in the {\it SHMEM} library of the UNICOS
operating system on the CRAY T3E and  Sgi Origin computing systems.\\
\fly is the result of the development of a preliminary software called WD99. This code
was developed in the period 1996-2000 using  several platforms. The first release was
developed with the CRAFT programming environment embedded the Cray T3D. The porting of the code on the Cray
T3E system, where the CRAFT was no more available, was performed in 1998 using the shmem library, that was the 
only one-side communication system available. The performances of the shmem library, in terms of scalability and
latency time are very good, being this library designed for the hardware architecture of the Cray T3E  and
of the Sgi Origin systems.
On systems like the \textsf{\large IBM SP} where these libraries are not available \fly has been modified
to use the local libraries.\\
The MPI-2 library was made available with good declared performances for IBM SP System only recently.
This library probably will be adopted in the next \fly version, to increase the code portability.\\
A more detailed treatment of the parallel computing techniques which have
been adopted and of the resulting performances on different systems can be found elsewhere.\\
Being an open source project, \fly can and must be modified to suit the particular needs of individual users.
In order to ease this, we will describe some features concerning the integration of the equations of motion
(section 2), the generation of initial conditions, the structure of the checkpoint
and the output  files.
In section 5  we  will also describe the structure of some parameter files needed 
to run a simulation. More detailed information
concerning the preparation of these parameter files, the compilation and running of the code can 
be found
in the \fly  User Guide \cite{flyug}.

\section{Equations of motion.}
The actual form of the discretized equations of motion is implemented in the subroutines
{\it upd\_pos} and {\it upd\_vel}; if the user needs to identify some
special parameters which are used in these subroutines, he may want to define them in
{\it initpars}. In the publicly distributed version of \fly, we have
implemented a set of cosmological equations of motion, solving the standard particle
equations of motion for a Friedmann cosmology. In the following, we will present these
equations and their discrete implementation.\\
The Friedmann-Robertson-Walker metric is characterized by an expansion factor $a(t)$, where $t$ is the conformal
time. Let $\boldsymbol{x}_{i}(t)$ be the comoving coordinate of the $i$-th particle and $m_{i}$ its mass, then the
equations of motion are given by:
\be 
\dot{\boldsymbol{x}_{i}} = \boldsymbol{v}_{i}  \label{eq1}
\ee 
\be
\dot{\boldsymbol{v}_{i}} + 2\frac{\dot{a}}{a}\boldsymbol{v}_{i} = - \frac{G}{a^{3}}
\sum_{j\ne i}\frac{m_{j}(\boldsymbol{x}_{i}-\boldsymbol{x}_{j})}
{\mid\boldsymbol{x}_{i}-\boldsymbol{x}_{j}\mid^{3}} +
\boldsymbol{F}_{Ewald}(\boldsymbol{x})  \label{eq2}
\ee
Here and in the following a dot denotes derivation w.r.t. the conformal time variable $t$, $G$ is the gravitational constant and the
last term, $\boldsymbol{F}_{Ewald}(\boldsymbol{x})$,  is the {\em Ewald correction},
which takes into account the contribution to the force from the infinite replicas of the
simulation box over the spatial directions. We also define the {\em Hubble constant}: 
$H(t)=\dot{a}/a$.\\
It is more convenient to introduce a set of dimensionless spatial, temporal and mass
variables:
\[
\boldsymbol{x}_{i}'=L_{0}\boldsymbol{x}_{i},\hspace*{0.5cm} t=t_{0}\tau,\hspace*{0.5cm}
m_{i}'=M_{0}m_{i}  
\]
In terms of these variables, the dimensionless equations of motion become:
\be
\frac{d\boldsymbol{x}_{i}'}{d\tau}=\boldsymbol{v}_{i}'  \label{eq3}
\ee
\be
\frac{d\boldsymbol{v}_{i}'}{d\tau}+2H(\tau)\boldsymbol{v}_{i}'=\frac{GM_{0}t_{0}^{2}}{L_{0}^{3}}
\frac{\boldsymbol{a}_{i}'}{a(\tau)^{3}}  \label{eq4}
\ee
The dimensionless comoving peculiar velocity $\boldsymbol{v}_{i}'$ is simply related to
the actual peculiar velocity by a scaling relation: $\boldsymbol{v}_{i} =
(L_{0}/t_{0})\boldsymbol{v}_{i}'$ (trivially deduced from eq.~\ref{eq3}, and the Hubble
constant $H(\tau)$ in dimensionless time units is given by:
\[
H(\tau)=t_{0}H(t)
\]
In fact, the latter relationship is easily obtained after having performed a change of
time variable in the definition of the Hubble constant:
\[
H(t)=\frac{1}{a}\frac{da}{dt}=\frac{1}{a}\frac{da}{d\tau}\frac{d
\tau}{dt}=\frac{1}{at_{0}}\frac{da}{d\tau}\equiv\frac{H(\tau)}
{t_{0}}
\]
From now on we will omit the apex from the dimensionless quantities.

\subsection{Units}
We choose units such that:
\be
\frac{GM_{0}t_{0}^{2}}{L_{0}^{3}}=1   \label{eq5}
\ee
so that the r.h.s. of eq.~\ref{eq4} has a unit multiplying factor. Apart from the  obvious reasons
of simplicity, this choice helps in reducing the number of floating-point operations in the
actual numerical implementation.\\
We have still to make choices for two out of the three the units entering eq.~\ref{eq5}
($L_{0}$, $t_{0}$ or $M_{0}$). We will adopt the values given by \cite{PDBook} for the
fundamental constants, rounded to the 6th decimal digit (but in the calculations we use up to
the 10th digit when available).\\
First, note that:
\[
GM_{\odot}=1.327125\times 10^{11} \frac{{\rm Km}^{3}}{{\rm
sec}^{2}}
\]
So, if we measure mass in units of $M_{\odot}$ we can rewrite:
$GM_{0}=GM_{\odot}(M_{0}/M_{\odot})$. From now on we will simply
write $M_{0}$ in place of $M_{0}/M_{\odot}$.\\
If we adopt as units of length:
\be
L_{0}=1\,{\rm Mpc}\, h^{-1}=3.08568\times 10^{19}\, h^{-1} {\rm Km}     \label{eq6}
\ee
and of time:
\be
t_{0}=2/3H_{0}=2.05759\times 10^{17} h^{-1}{\rm sec}        \label{eq7}
\ee
we get from eq.~\ref{eq5} the units of mass:
\be
M_{0}=\frac{L_{0}^{3}}{t_{0}^{2}GM_{\odot}}=5.22904\times
10^{12}h^{-1} M_{\odot}      \label{eq8}
\ee
These are the fundamental units which we have implicitly adopted in our discretized equations.
We can now easily deduce the units of other relevant quantities:
\begin{itemize}
\item {\bf Velocity}:
\be
{\rm v}_{0}=\frac{L_{0}}{t_{0}}=146.21794\, \frac{{\rm Km}}{{\rm sec}}
\ee
\item {\bf Potential} {\large \em (per unit mass)}:
\be
\phi_{0} = \left(\frac{L_{0}}{t_{0}}\right)^{2}=2.137971\times 10^{4}
\left(\frac{{\rm Km}}{{\rm sec}}\right)^{2}
\ee
\end{itemize}
Finally, we can deduce the values of some useful quantities in these units:
\begin{itemize}
\item {\bf Gravitational constant}. -
Using eqs.~\ref{eq5}-\ref{eq8}, we obtain:
\be
{\rm G}=\frac{L_{0}^{3}}{M_{0}t_{0}^{2}} = \frac{v_{0}^{2}L_{0}}{M_{0}}
\ee
\item {\bf Hubble constant}. - Using the units for velocity we
get:
\be
{\rm H}_{0}=100 \frac{{\rm Km}}{{\rm sec}\cdot
{\rm Mpc}}=\frac{100}{146.218}\frac{v_{0}}{L_{0}}=0.68391
\frac{v_{0}}{L_{0}}
\label{eq:h0}
\ee
\item {\bf Critical Density}. - Using the definitions above we
get:
\be
\rho_{c}=\frac{3{\rm H}_{0}^{2}}{8\pi{\rm G}}h^{2}=5.583\times 10^{-2}h^{2}
\ee
\end{itemize}

\subsection{Choice of the time variable}
In eqs.~\ref{eq3}-\ref{eq4} the dimensionless time $\tau$ is used as time variable. In actual calculations
this is not necessarily the best choice, because often the actual time variable adopted in the outputs is
the redshift, which bears a nonlinear relationship with the conformal time, which is also dependent on the
underlying cosmological model. We have then adopted a new time variable, introduced first  in \cite{1985ApJS...57..241E}:
\[
p=a^{\alpha}
\]
where $\alpha $ is a coefficient to be determined by the user. For closed
models ($\Omega_{0}=1$) G. Efstathiou \cite{1985ApJS...57..241E} advise a value $\alpha = 1/2$, which
ensures an approximately constant r.m.s. advancement of particle positions from one time-step to the next. We choose a constant step $\Delta p$ in order to secure that the leapfrog intergrator keeps second-order accuracy throughout the run. \\
With this change of variable the new equations of motion are:
\be
\frac{d\boldsymbol{x}_{i}}{dp}=\boldsymbol{u}_{i}  \label{eq14}
\ee
\be
\frac{d\boldsymbol{u}_{i}}{dp}+2A(p)\boldsymbol{u}_{i}=B(p)
{\boldsymbol{a}_{i}}  \label{eq15}
\ee
where the coefficients are given by:
\be
A(p) = \frac{1 + \alpha + \ddot{a}a/\dot{a}^{2}}{2\alpha a^{\alpha}}  \label{eq16}
\ee
\be
B(p) = \frac{1}{\alpha^{2}\dot{a}^{2} a^{2\alpha + 1}}  \label{eq17}
\ee
 (cf. eqs. 10a-b of G. Efstathiou \cite{1985ApJS...57..241E}). The numerical evaluation of the time
derivatives $\dot{a}, \ddot{a}$ is rather cumbersome, but it can be simplified by using some
combinations of the Friedmann equations. From  S. Carrol (2000, eq. 32) \cite{2000astro-ph...0004075} we have:
\be
\ddot{a}a/\dot{a}^{2} = \Omega_{\Lambda} - \frac{1}{2}\Omega_{m}     \label{eq18}
\ee
Moreover, one can easily verify (see e.g. \cite{1997astro-ph...9712217}) that one of the Friedmann
equations can be written as:
\be
\dot{a}^{2}a = H^{2}\left[\Omega_{m}(z=0) + \Omega_{k}a + \Omega_{\Lambda}(z=0)a^{3} \right]
\label{eq19}
\ee
Here we have introduced the standard definitions for the dimensionless baryonic ($\Omega_{m}$),
curvature ($\Omega_{k}$) and vacuum ($\Omega_{\Lambda}$) parameters:
\[
\Omega_{m} = \frac{8\pi{\rm G}\rho}{3H^{2}}, \hspace*{1.5cm}\Omega_{\Lambda} = \frac{\Lambda}{3H^{2}}
\]
Substituting eqs.~\ref{eq18}-\ref{eq19} in eqs.~\ref{eq16} and~\ref{eq17}, respectively, we
get:
\be
A(p) = \frac{1 + \alpha + \Omega_{\Lambda} - \frac{1}{2}\Omega_{m} }{2\alpha p}  \label{eq20}
\ee
\be
B(p) = \frac{1}{\alpha^{2}H^{2}p^{2}\left[\Omega_{m} + \Omega_{\Lambda}a^{3} \right]}
\label{eq21}
\ee
Note that in the equations above the cosmological parameters $\Omega_{m},  \Omega_{\Lambda}$ are all evaluated at $z=0$.
The numerical evaluation of eqs.~\ref{eq20} and \ref{eq21} is simpler than that of the original
definitions. Some of the factors appearing in these terms are in fact constant and are
computed only once in the initialization subroutine {\it inparams}.

\subsection{Remark about initial velocities}
As is clear from the preceding paragraph, the velocity $\boldsymbol{u}$ does not coincide with the dimensionless peculiar velocity 
$\boldsymbol{v}'$. One must keep this fact in mind when  computing the initial positions and velocities. From the definition of peculiar velocity in dimensionless units (eq.~\ref{eq3}), using the parameter $p$ given above, and also eq.~\ref{eq14} we get:
\[
\boldsymbol{v}_{i}'=\frac{d\boldsymbol{x}_{i}}{d\tau}=\frac{d\boldsymbol{x}_{i}}{dp}\frac{dp}{dt}t_{0}=\boldsymbol{u}\alpha\left[ a(t)\right]^{\alpha-1}\dot{a}\frac{2}{3H_{0}} =
\]
\[
= \boldsymbol{u}\alpha \frac{p}{a} \cdot \frac{2}{3}a\left[\frac{\Omega_{m}}{a^{3}} + \Omega_{\Lambda}\right]^{1/2}
\]
Finally, in terms of the expansion factor at the start of the simulation $a_{in}=(1+z_{in})^{-1}$, we obtain:
\be
\boldsymbol{u}_{i} = \frac{3}{2\alpha} a_{i}^{-\alpha}\left[\frac{\Omega_{m}}{a_{i}^{3}} + \Omega_{\Lambda}\right]^{-1/2} \frac{\boldsymbol{v}_{i}}{v_{0}}
\ee
where $\boldsymbol{v}_{i}$ is the peculiar velocity in dimensional units as given for instance, by the COSMICS code \cite{1995astro-ph...9506070}

\subsection{Gravitational potential}
The gravitational potential adopted to compute the acceleration is given by the Plummer form
\be
\Phi(r) = - \frac {{\rm G} m_1 m_2}{(r^2+\epsilon^2)^{1/2}}
\ee

where $\epsilon$ is a softening length. For $\epsilon > 0$
this potential is finite at the origin, and is adopted in order to avoid the formation of tight binary pairs.
 
\subsection{Discretized equations}
It is customary to adopt a leapfrog discretization scheme for N-body cosmological simulations \cite{1985ApJS...57..241E},
 so that our final equations of motion are:
\be
\boldsymbol{u}_{i}^{n+1/2} = \boldsymbol{u}_{i}^{n-1/2}
\frac{1 - A(p_{n})\Delta p}{1 + A(p_{n})\Delta p}  +
\frac{B(p_{n})\Delta p}{1 + A(p_{n})\Delta p}\boldsymbol{a}_{i}^{n}
\label{eq22}
\ee
\be
\boldsymbol{x}_{i}^{n+1} = \boldsymbol{x}_{i}^{n} + \boldsymbol{u}_{i}^{n+1/2}\Delta p
\label{eq23}
\ee
where $\boldsymbol{a}_{i}^{n}$ is the acceleration, the lower index refers to the particle
($1\leq i \leq N_{bodies}$) and the upper one to the time-step. Note that positions and
velocities are evaluated at steps differing by one-half time-step: this must be taken into
account when preparating the initial conditions. 
Actually before  starting the run, the subroutine {\it leap\_corr} executes this phase adjustment by advancing the position of half time-step, using a simple second order extrapolation of particles' positions.

\subsection{Choice of the time-step}
Also the choice of the time stepping criterion should be left to the user's own choice. In \fly the time stepping criterion is performed
in the subroutine {\it dt\_comp}. As a default, we adopt a criterion based on the evaluation of the maximum acceleration among all the particles:
\be
\Delta p = \left[\frac{\alpha\epsilon_{soft}}{{\rm max}_{i=1, N}\mid\boldsymbol{a}_{i}\mid}\right]
\label{eq:dp}
\ee
where $\epsilon_{soft}, {\rm max}_{i=1, N}\mid\boldsymbol{a}_{i}\mid$ are the softening length and the maximum acceleration, and
$\alpha\leq 1$ is a parameter which can be freely adjusted. This time stepping criterion is dynamic, and guarantees an almost constant
r.m.s. advancement of the particle with a large velocity. However, the final choice is obviously left to the user, who can easily modify
the subroutine {\it dt\_comp} or can give a constant time-step (see par. 5.4).

\section{The FLY grouping}
A tree code computes the force on a particle, by means of the interaction between the particle, and
some elements (body or cells) of the tree that forms the interaction list $IL_p$ for each particle.
In a classical tree schema each particle must have its own $IL_p$ to compute the force on it.\\
The fundamental idea of the tree codes consists in  the approximation 
of the force component for a particle. Considering a region $\gamma$, the force component 
on an {\it i-th} particle may be computed as

\begin{equation}
\sum_{j\in\gamma} - \frac{Gm_j {\bf d}_{ij}}{\mid d_{ij}\mid^3} \approx \frac{GM {\bf d}_{i,cm}}
{\mid d_{i,cm}\mid^3} + \; \hbox{higher order multipoles terms }\; 
\label{eq:mu}
\end{equation}
 where $M = \sum_{j\in\gamma}m_j$ and $cm$ is the  center of mass of $\gamma$.\\
In  eq. (\ref{eq:mu}) the multipole expansion is carried out up to the quadrupole order when a {\it far} group is 
considered.
The tree  method,  having no geometrical constraint, adapts dynamically the tree structure to the 
particles distribution  and to the clusters, without loss of accuracy. This method scales as $O(NlogN)$.\\
 The criterion for 
determining whether a cell must be included in the  $IL_p$
 is based on an opening angle 
parameter $ \theta$  given by
\begin{equation}
\frac{C_l}{d} \le \theta ,
\label{eq:un}
\end{equation}
 where $C_l$ is the size of the cell and $d$ is the distance of 
$ p$ from the center of mass of the cell. The smaller the values of $ \theta$, the more the number of cells 
to be opened and hence the more accuracy
of forces (for $ \theta = 0.8$ we have an rms error lower than 1\% on the accelerations using a multipole
expansion of the quadrupole order \cite{her87}).\\
\fly uses a grouping method to form a grouped  interaction list $IL_g$, assigned to all the particles 
lying in grouping cell $C_{group}$.\\ 
The same above mentioned method is applied to a hypothetical particle 
placed in the center of mass of the $C_{group}$, hereafter VB (Virtual Body). Moreover, we consider the
 $IL_g$ as formed by 
two parts given by
\begin{equation} 
IL_g = IL_{far} + IL_{near}
\label{eq:il_g}
\end{equation}
$ IL_{far}$ and $IL_{near}$ being two subsets of the interaction list. 
An element is included in one of the two subsets, using the following Sphere method for all the elements that satisfy eq. (\ref{eq:un}).

\vspace*{1.0cm}

\noindent \hspace*{1.0cm}{\bf Define} $Sphere_{radius} = 3  \frac {Cellsize(C_{group}) \sqrt{3}}{2}$\\
\\
\\
\vspace*{0.3cm}
\hspace*{1.0cm}{\bf If} $Distance(IL_g(element),VB) >  Sphere_{radius}$ \\
\vspace*{0.3cm}
\hspace*{2.0cm} {\it Add element to} $IL_{far}$\\
\vspace*{0.3cm}
\hspace*{1.0cm}{\bf Else}\\
\vspace*{0.3cm}
\hspace*{2.0cm}   {\it Add element to} $IL_{near}$\\ 
\vspace*{0.3cm}
\hspace*{1.0cm}{\bf Endif}\\
\vspace*{1.0cm}

\noindent Moreover all the particles $p \in C_{group}$ are included in $IL_{near}$.\\
Using the two subsets it is possible to compute the  force ${\bf F}_p$ 
on a particle $ p \in C_{group}$ as the sum of two components,
\begin{equation} 
{\bf F}_p ={\bf F}_{far} +{\bf F}_{near} ,
\label{eq:F_p}
\end{equation}

\noindent where ${\bf F}_{far}$ is a  force component due to 
the elements listed in $IL_{far}$ and ${\bf F}_{near}$ is the force component
due to the elements in $IL_{near}$. 
We assume the  component  ${\bf F}_{far}$  
to be the same for each particle $p \in C_{group}$  and compute it considering the gravitational interaction between the VB
  and only the elements listed in 
$IL_{far}$, 
while the  ${\bf F}_{near}$ component is computed separately for each $p$ 
particle  by the direct interaction with the elements listed in 
$IL_{near}$. 

\subsection{Error Analysis}
The dominant component of the error of the classical Barnes-Hut method arise from the truncated multipole series of degree p (generally
$p=2$) in eq \ref{eq:mu}, and is O(${\gamma}^{p+1}$) where here $\gamma$ is the inverse of $\theta$ (eq. \ref{eq:un}) and
is always greather than 1. The rms error, in a typical Large scale Structure (LSS) simulation with $\theta=0.8$ is
about equal to $1\%$ on the acceleration, in comparison with a direct particle-particle method
\cite{1989ApJS...70..389B}.\\ 
The error we introduce with the \fly grouping method depends on the choice of some parameters.
The "LIV. GROU" parameter (see par. 5.5) is the level of the tree where a group can be formed. This parameter
sets the maximum $C_{group}$ size that must be chosen in order to ensure that the difference between $IL_p$ and $IL_{VB}$ is 
negligible (no more than $1\%$ of the elements). The "BODY GROU" parameter (see par. 5.5) sets the maximum 
number of particles that can form a group within a cell: all these particles are listed in $IL_{near}$ and there is a
direct particle-particle interaction among these elements that form a group.\\
For a LSS simulation in a 50 Mpc box  with more than 2
million  particles the "LIV. GROU" parameter can be set to the sixth level of the tree and the "BODY GROU" 
parameter equal to 32. These values maintain the global error around to $1\%$ and good performances of the
code.
Further detail on this method, the errors and the performance of \fly can be found in  \cite{bec2001}.

\section{Program organisation}
This section list all the main subroutines and functions of \fly. The code is organized in some subroutines
that are executed  at the start of the job only and other subroutines that are executed for each time-step.

\subsection{Main and module}
{\bf fly\_h} is the module file of \fly, where all the global variables are defined.\\
{\bf fly} is the main program of \fly, that starts the system inizialization and the execution of all 
the programmed time-step.\\

\subsection{System inizialization}
The following subroutines are executed before  the first time-step, and are listed in the same order
of execution during the start-up phase.\\
{\bf null} initializes some variables.\\
{\bf sys\_init} calls all the following subroutines to read parameters and initial condition files, at the start of the run.\\
{\bf read\_params} reads the input parameters from {\it stat\_pars}  and {\it dyn\_pars} files, at the start of the run.\\
{\bf read\_redsh} reads the redshift list that will be used to produce  the output files.\\
{\bf init\_ew} reads and sets the Ewald tables for periodical boundary conditions.\\
{\bf read\_b\_bin} reads the binary checkpoint file (i.e. initial condition file) with positions and velocities of all particles, at the start of the run.\\
{\bf read\_b\_asc} reads the ascii checkpoint file (i.e. initial condition file) with positions and velocities of all particles, at the start of the run.\\
{\bf init\_balance} distributes the initial load among the processors.\\
{\bf init\_pars} initializes  physical variables at the start of the run.\\
{\bf init\_pos} executes the Leapfrog correction on the input data, at the start of the simulation.\\
{\bf reset\_pos} resets all particles inside the box where the simulation will evolves, at the start of the simulation.\\

\subsection{Time-step execution}
The following subroutines are executed for all the time-step of a single job, at the end of the inizialization
phase. The number of steps executed by a job being the "NUM. STEP" value in the {\it stat\_pars} file (see par.
5.4). The subroutines are listed  in the same order of execution.\\
{\bf inpar\_dyn} reads the input parameters from {\it dyn\_pars} files, at the start of each time-step cycle.\\
{\bf step} executes a time-step cycle: it calls {\it step\_force}, {\it dt\_comp}, {\it upd\_vel}, {\it upd\_pos}, {\it out\_32} and {\it wr\_native}.\\
{\bf step\_force} builds the tree and computes the acceleration on the particles: it calls {\it tree\_build}, {\it acc\_comp}.\\
{\bf ch\_all} allocates  RAM for temporary remote data storage.\\
{\bf tree\_build} builds the tree data structure. It calls {\it tree\_gen}, {\it find\_group} and {\it cell\_prop}.\\
{\bf tree\_gen} builds the geometry of the tree.\\
{\bf find\_group} marks all the grouping cell (see sect. 3).\\
{\bf cell\_prop} computes the monopole and  the quadrupole momentum for all the tree cells.\\
{\bf acc\_comp} starts {\it ilist\_group}, {\it force\_group}, {\it ilist} and {\it force} subroutines.\\
{\bf ilist\_group} walks in the tree and forms the interaction list for each grouping cell.\\
{\bf force\_group} computes the grouping acceleration for all particles belonging to a grouping cell, using the interaction list formed
by ilist\_group. These components will be completed in the {\it force} subroutine.\\
{\bf ilist} walks in the tree and forms the interaction list, for each particle.\\
{\bf force} computes the acceleration on a particle using the interaction list formed by ilist, or complete the acceleration
components  for the particles belonging to a grouping cells.\\
{\bf dt\_comp} computes the  adaptive time-step (see par. 2.6).\\
{\bf upd\_vel} advances the particle velocities at the end of a time-step cycle.\\
{\bf upd\_pos} advances the particle positions at the end of a time-step cycle.\\
{\bf out\_32} produces the programmed output and the quick-look files using the table read in {\it read\_redsh}.\\
{\bf leapf\_corr} executes the Leapfrog correction on data output.\\
{\bf wr\_native} writes bodies on the checkpoint file: it calls {\it write\_b\_bin} or {\it write\_b\_asc}.\\
{\bf write\_b\_bin} writes the binary checkpoint file, with positions and velocities of all particles, at the end of the run.\\
{\bf write\_b\_asc} writes the ascii checkpoint file, with positions and velocities of all particles, at the end of the run.\\

At the end of each time-step the main program automatically computes the new parameters for the {\it Dynamic Load
Balance} and writes in the default output the timing of each processor during the step.

\section{Data input}
\subsection{Initial Conditions}
\fly does not give the program for the initial condition  generation, but it is possible to use any program 
the user likes (e.g. COSMICS by E. Bertschinger), provided that the input file is prepared as described here. 
The input file simply contains all the positions and velocities of all the particles in double precision 
C format (8 byte for each component). \fly uses a C routine to read and write the input/output particles data 
file. The initial file input must be produced using the C language, like the \fly io.c function:
\begin{verbatim}
void write_b_bin(double *pos,double *vel,char *cfilename,signed int *nlong)
{
  FILE *f_b_bin;

  f_b_bin = fopen(cfilename,"wb");
  fwrite(pos,sizeof(double),*nlong,f_b_bin);
  fwrite(vel,sizeof(double),*nlong,f_b_bin);
  fclose(f_b_bin);
}
\end{verbatim}
\noindent
Note that the {\bf pos} and {\bf vel} arrays should be defined as: {\bf pos[nbodies][3], vel[nbodies][3]}
in order to keep the same memory map distribution they have in Fortran 90, considering that
in the C language, the array ordering is the inverse as in Fortran. 
So the above ordering corresponds to the following Fortran ordering:
{\bf pos(3,1:nbodies), vel(3,1:nbodies)}.\\
It can often be advisable to create a  /temp/FLY directory (where the system allows a great capability
 area on the /temp file system) to put the initial condition file.\\
On Cray T3E and Sgi Origin if you use a Fortran program to produce the initial condition file, it should be produced {\em without control characters}, i.e. it must have an  unblocked file structure with "stdio" style buffering, compatible with the C {\em fwrite} and {\em fread} functions. The following code excerpt gives an example:
\begin{verbatim}
	REAL(KIND=8),	DIMENSION(3,1:nbodies) :: pos,vel 
 
	....generate the  pos and vel arrays... 
	 
	OPEN(UNIT=12,FILE='posvel_0',FORM='UNFORMATTED',STATUS='NEW') 
 
	write(12) pos 
	write(12) vel 
		    
	CLOSE(UNIT=12) 
\end{verbatim}

\noindent
But in the Cray T3E and the Sgi Origin systems you should give the following command at the system prompt, to generate an  unblocked file structure with "stdio" style buffering:
\begin{verbatim}
prompt>  assign -s sbin u:12
\end{verbatim}
Otherwise an ASCII input file can be supplied (for all platforms)  with the following data format:
\begin{verbatim}
	REAL(KIND=8),	DIMENSION(3,1:nbodies) :: pos,vel 
 
	....generate the  pos and vel arrays... 
	 
	OPEN(UNIT=12,FILE='posvel_0',STATUS='NEW') 
 
	DO i=1,nbodies 
		write(12,1000) pos(1:3,i) 
	ENDDO 
 	DO i=1,nbodies 
		write(12,1000) vel(1:3,i) 
	ENDDO 
     
	CLOSE(UNIT=12) 
 
1000	FORMAT(3(1X,F20.10))
\end{verbatim}

\subsection{Parameter files}
You can use the graphical interface to generate all input parameters files, if in the system you use 
has the Tcl/Tk is installed. The commands
\begin{verbatim}
prompt> cd ./FLY_2.1/src/tcl 
prompt> wish fly_2.1.tcl 
\end{verbatim}

start the graphical interface that will guide the user to create the parameter files and the Makefile, 
to compile the code and  
to create the platform-dependent scripts which allow the user to submit a job in the system queue.\\
We also provide the user with an {\em assistant} program to create the  parameter
files and to compile \fly. The assistant is 
invoked using the command  {\bf assistant\_t3e} on Cray T3E ({\bf assistant\_ori} or {\bf assistant\_sp3} on Sgi Origin and IBM SP, 
respectively) that the user will find in the directory ./FLY\_2.1/src: 
\begin{verbatim}
prompt> cd ./FLY_2.1/src 
prompt> ./assistant_t3e 
\end{verbatim}

The assistant program is structured with several sections and, where applicable, it gives a default value. 
The user must set a {\em Working Directory}  and an {\em Executable Directory} to store 
the parameter files and the executable programs respectively. 
The assistant does not create these directories, and the user must create them before launching the assistant.\\
After having specified these directories, the {\em assistant} will create the input parameter files, and the specific
module file that will be used during the compilation.

\subsection{Generating the ./FLY\_2.1/bin/fly\_fnames file}
This file contains the filenames (including the path) of all the input parameter files that will be used 
during the run. 
If this file does not exist, the {\it assistant} program will generate it and ask for the filenames: the 
default filenames are reported in square brackets:
\begin{verbatim}
Starting generation of ../bin/fly_fnames file...                    
 
Do you want to create this file (Y/N) : y
 
 
All NOT-ABSOLUTE filename paths will be ../bin/ : 
 
Please supply stat_pars filename [stat_pars] : 
 
Please supply dyn_pars filename [dyn_pars] : 
 
Please supply ew_grid filename [ew_grid] : 
 
Please supply ew_tab filename [ew_tab] : 
 
Please supply out32.tab filename [out32.tab] : 
 
Please supply ql.tab filename [ql.tab] :
\end{verbatim}

The file will be created, containing the filenames the user gives. In the following we will suppose the user has chosen the default filenames.
A more detailed discussion about all the files and the parameters can be found in the \fly User Guide.

\subsection{The stat\_pars file}
This file has the following structure:
\begin{verbatim}
HEADER   =try_sim                     
NUM. STEP=  10
MAX_TIME =  0
BAL. PAR.= 0.90
DELTA T. =   0.00450
DT VAR.  =T
OPEN PAR.= 0.90
SOFT PAR.=   0.01000
QUADRUP. =T
OMEGA.CDM=   0.30000
OMEGA.HDM=   0.30000
OMEGA.LAM=   0.40000
HUB_CONST=  0.650000
N. BODIES=     2097152
MASS BODY=   0.00120
IBOD_FILE=/tmp/FLY/posvel_ 
IBOD_TYPE=B            
OBOD_FILE=/tmp/FLY/out_                 
OBOD_TYPE=B            
QLK_FILE =/tmp/FLY/qlk_                 
\end{verbatim}
This file contains some data on the simulations that must be never changed during the system evolution. 
The assistant  asks for each parameter and, eventually, gives a default value. Some peculiar parameters are:
\begin{itemize}
\item{HEADER is a string to identify the simulation.} 
\item{NUM. STEP is the number of time-step cycles of each job.} 
\item{MAX\_TIME is the maximum CPU time allowed for each job.  This value must be lower  than or 
equal to the single CPU time as specified in
the script job. If this value is greater than 0 the NUM. STEP parameter is negligible  and the number 
of time-step cycles for a single job, is automatically computed by \fly.}
\item{BAL. PAR. is used to balance the load among  the processors at the start of a run. This value is 
automatically computed at the end of each time-step cycle, to allow a perfectly load balance. 
It is the percentage of the local particles that  must be computed by the local processor, the remaining 
portion being computed by the first available processor. 
We recommend to use the default value 0.90.}
\item{DELTA T. is the integrator time-step used in the system evolution . For cosmological simulations a
 possible choice can be found in  \cite{1985ApJS...57..241E}.}
\item{DT VAR. \fly can also use an adaptive Delta T integrator   automatically computed in dt\_comp.F subroutine
(see par. 2.6). In this case the user must give the T (True) value to this parameter. The F (False) value, forces
\fly to use a fixed Delta T as given above.}
\item{OPEN PAR is the opening angle parameter ($\theta$), used to open or close the cells during the tree 
walking phase, to build the interaction list for each particle (eq.\ref{eq:un}).}
\item{SOFT PAR.  is the softening length of the gravitational interaction (see par. 2.4).}
\item{IBOD\_FILE  This is the root filename of the checkpoint file, to which a counter 
is automatically added  (as a suffix) to form the complete filename.}
\item{IBOD\_TYPE. The user must specify B for binary or A for ascii data format for the IBOD\_FILE data type.}
\item{OBOD\_FILE This is the root filename of the output files, to which a counter, giving the redshift value, is
automatically added to form the complete filename. These output files are the scientific data output produced by \fly with the Leapfrog 
correction for the phase adjustment. The data format of these files is the same as the checkpoint files.}
\item{OBOD\_TYPE. The user must specify B for binary or A for ascii data format for the OBOD\_FILE data type.}
\item{QLK\_FILE This is the root filename of the quick-look files, to which a counter, giving the redshift value, is
automatically added to form the complete filename. This is the ASCII file  containing random particles  positions
and velocity( see par.5.6). During the system simulation a header file (i.e. /tmp/FLY/qlk\_hea) will be
 automatically created. This file can be used to visualized the system evolution using this file as the header file of AstroMD (http://www.cineca.it/astromd) 
 
The qlk\_hea file has a format like the following:
\begin{verbatim}
header_line                            
ASCII
time
262144   
17  /gpfs/temp/ube/data262k/qlk_60.0000
10  /gpfs/temp/ube/data262k/qlk_50.0000
68  /gpfs/temp/ube/data262k/qlk_20.0000
\end{verbatim}
The value 262144 is the number of data points included in the quick-look file, the filename suffix ( \_60.0000) 
is the corresponding redshift value when the output 
is produced, and the number before the filenames (17, 10, 68) 
is the number of time-step cycles that \fly has reined from the start or the previous quick-look file produced.} 

\end{itemize}

\subsection{The dyn\_pars file}
This file contains some parameters of a simulation, that could be changed during the system evolution, and has the following structure:
\begin{verbatim}
CURR.TIME=  50.000000000000000
CURR.STEP=  0
MAX  STEP=100
LIV. GROU=  9
BODY GROU= 16
GROUP FL.=  1
SORT_LEV.=  3
BOX COMP.=  0
BOX SIZE =  50.0000
X MIN VER=  0.000000
Y MIN VER=  0.000000
Z MIN VER=  0.000000
\end{verbatim}
This file must be created by the user and is read at the beginning of each time-step cycle during a run. This file is automatically updated  at the end of each job. 
The assistant  asks for each parameter and, eventually, gives a default value. Some peculiar parameter are:
\begin{itemize}
\item{CURR.TIME  is the redshift value at the current time-step cycle.} 
 
\item{CURR.STEP. is the current time-step cycle. We strongly recommend to start from 0 as initial time-step cycle}
\item{MAX  STEP. is the maximum allowable time-step cycle.  \fly halts the simulation when this value, or the
final redshift given by the user (see par. 5.6) is reached.}

\item{LIV. GROU is the level where the grouping cell can be started to form, being LIV. GROU=0 the root level .} 
\item{BODY GROU is the maximum number of particles $N_{crit}$ that a grouping cell can contain (see above).
We recommend to use a value from 8 up to 32.}
\item{SORT LEV. is automatically computed by the assistant, and is used by the FLY\_sort utility, that
builds  a tree up to the level indicated by this parameter, and produces a sorted data input file, organized as the
tree structure} 
\end{itemize}
The items GROUP FL. and BOX COMP are for future usage, and the items BOX SIZE, X-Y-Z MIN VER are respectively the size of the box and the 
coordinates of the lowest vertex.

\subsection{Other files}
The following files must be located in the working directory of \fly and contain the following data.

\begin{itemize}
\item{out32.tab. List of programmed redshift outputs, \fly generates the output and the quick-look files, with
the root filename reported in the  
OBOD\_FILE and QLK\_FILE parameters of the stat\_pars file,  and a suffix given by the redshift value listed in this table.}
\item{ql.tab is a random  sequence of values used to generate quick-look file.}
\item{ew\_grid and ew\_tab. These tables are used for  the {\em Ewald correction} considering the boundary periodical conditions.}

\end{itemize}
More details on these files can be found in the \fly User Guide.

\section{Installing FLY}
The \fly installation procedure is straightforward and  even the most unexperienced reader can compile and run successfully a 
parallel code. \\
Download the file FLY.tar.gz and copy it somewhere (e.g. in your HOME directory)  of a Cray T3E
system and/or Sgi Origin  and/or an IBM  SP system.\\
Make   sure  that  a subdirectory with the same name ./FLY\_2.1 is not present, to avoid possible conflicts.

\subsection{Unpacking FLY}
Once you have downloaded the package, give the following commands to uncompress and unfold it:\\  \\
\$ gunzip FLY.tar.gz\\
\$ tar -xvf FLY.tar\\ \\

The tar command  will create the following directories:

./FLY\_2.1			(FLY installation directory)\\
./FLY\_2.1/bin		(Executable programs and input parameters files) \\
./FLY\_2.1/src		(Source code and assistant program)   \\
./FLY\_2.1/src/tcl	(TCL program: graphical interface)      \\
./FLY\_2.1/job		(Job script to run FLY)             \\
./FLY\_2.1/out\_stat	(Output log file of the FLY run)   \\
./FLY\_2.1/out\_log	(Output log file of the job script)   \\
./FLY\_2.1/Testcase	(Input/output example files)   \\
./FLY\_2.1/tools        (Utilities) \\ 
./FLY\_2.1/FLY\_ug.doc   (\fly User Guide) \\ 

Now you can use the graphical interface to generate all the parameters files and the executable
program (see par. 5.2). 

\subsection{FLY module, makefile and script files}
The fly\_h.F file is a module file containing all the fundamental data that are 
common to all the subroutines of \fly.
To generate an executable program, the user must set some fundamental parameters.\\
The user must give the number of processors of the parallel execution, the number of particles that he wants to use (the same as stat\_pars file),
and the estimated ram available for each processor (this value is used to allocate dynamically a temporary buffer in the local ram),
and the maximum length of the interaction list $IL$, formed during the tree walk procedure: this value
 depends on the box size, the number of bodies, and the opening parameter: in a clustered regions this value 
is greater than in an uniform region. 
It is very important, for the \fly performance, to give a  safety value but close to the expected maximum length of the interaction list.\\
All the parameters have a default value that is set taking a standard cosmological model as a template.\\ 
The mkfl\_fly file is the makefile of \fly . The user must give the  filename of the executable program and 
the executable directory, the user can also specify if he wants receive a statistical report. 
The statistical report concerns only the code performance, it  must be used only from expert users of \fly,
and  must not be used during a production run. \fly also has the mkfl\_fly\_sort file to generate the FLY\_sort utility.\\ 
The last file the user can create is the FLY\_job.cmd. This is a script file that contains a schema to 
submit a job to the system queue.  \fly also creates the FLY\_sort\_job.cmd to submit the FLY\_sort  utility.
On IBM SP, some information about the node characteristics and the system topology are also asked to generate
the script file. In any case the user must customize both the FLY\_job.cmd and FLY\_sort\_job.cmd files.
\subsubsection{Compiler options}
FLY automatically sets the compiler options, that can be changed by expert users. All the options were well
tested in order to reach the best performance of the code. The following options are given:\\

{\bf Cray T3E system}
\begin{itemize}
\item{compilers: f90 and cc}
\item{{\bf -N132} : specify the line width to 132 column line;}
\item{{\bf -O3} : aggressive optimization;}
\item{{\bf -LANG:recursive=on} : the compiler assumes that a statically allocated local variable could be
referenced or modified by a recursive procedure call;}
\item{{\bf -DT3E} : must be always given. The T3E flag is used to compile specific section of the code for this
specific platform;}
\item{{\bf -DPW2} : must be used to obtain higher performances when the number of processors that will be used
to run a simulation is a power of two;}
\item{{\bf -DSTA} : is used to obtain a statistical report on the number of remote GET and PUT calls. can be
used to test the fly efficiency and must not be used during the simulation run;}
\item{{\bf -DSORT} : generate the \fly\_sort utility.}
\end{itemize}

{\bf SGI Origin system}
\begin{itemize}
\item{compilers: f90 and cc}
\item{{\bf -64} : generate a 64 bit objects;}
\item{{\bf -extended\_source} : specify the line width to 132 column line;}
\item{{\bf -O3} : see Cray T3E system option;}
\item{{\bf -LANG:recursive=on} : see Cray T3E system option;}
\item{{\bf -DORIGIN, -DPW2, -DSTA, -DSORT} : see Cray T3E system options;}
\end{itemize}

{\bf IBM SP system}
\begin{itemize}
\item{compilers: mpxlf90\_r and xlc}
\item{{\bf -O4} : aggressive optimization;}
\item{{\bf -qrealsize=8} : sets default real to 8 bytes;}
\item{{\bf -qfloat=fltint:rsqrt} : optimizations for floating point operations;}
\item{{\bf -qmaxmem=-1} : unlimited memory  used by space intensive optimizations;}
\item{{\bf -DSP3, -DPW2, -DSTA, -DSORT} : see Cray T3E system options;}
\end{itemize}

\section{Starting a simulation} 
The parameters that the user must set to start a cosmological simulation, are listed in the 
{\it stat\_pars} file (see par. 5.4). The most important parameters 
that describe the cosmological model and the time evolution of a simulation are listed in the  
following. 
\begin{itemize} 
{\item {\bf DELTA T.} Is the integrator time-step used in the system evolution. \fly uses the 
Leapfrog second order integrator for the system dynamic. The user can set a constant safe  
value to ensure the accuracy, the stability and the efficiency of the run. This value is ignored  
if the user want to use the adaptive integrator time-step, included in the \fly code.} 
{\item {\bf DT VAR.} This is a logical value. A True value (T) means that the user want use 
an adaptive integrator time-step. \fly computes each time-step the integrator value as described in the 
eq. \ref{eq:dp} }  
{\item {\bf OPEN PAR.} This value is the $\theta$ opening angle parameter that is used in the Barnes-Hut 
algorithm. A discussion of this parameter is given in the sect. 3 eq. \ref{eq:un}. Typical values 
used for the LSS cosmological simulation are in the range 0.6 - 1.0. } 
{\item {\bf SOFT PAR} A softened force of the form 
\be 
\boldsymbol{F}=\frac{-GMm\boldsymbol{R}}{(R^2+\epsilon^2)^{3/2}}  
\label{eq:sof} 
\ee 
is adopted in the \fly code, where $\epsilon$ is the softening parameter.\\ 
This parameter is crucial in determining the amount of {\it hardening} of the gravitational interaction 
ultimately the amount of small scale substructure which is not destroyed during the gravitational 
evolution of the system.}  
{\item {\bf QUADRUP.} This is a logical value. A true value (T) means that the quadrupole  
order will be used in the eq. \ref{eq:mu}. A false value means that the higher order multipole terms 
of the force component will not be considered. In order to have a good accuracy for the force component,  
the true value is strongly recommended.} 
{\item {\bf OMEGA CDM}, {\bf OMEGA HDM} and {\bf OMEGA LAM} These parameters must be set by the scientist and 
are used to select a cosmological model for the simulation. The summation of all these values must 
be equal to 1.0.}  
{\item {\bf HUB\_CONST} Is the Hubble constant $H_0$ (eq. \ref{eq:h0}) express in the unit of  
$100 Km/sec/Mpc$. A value of 0.65 means $H_0=65 Km/sec/Mpc$.}  
{\item {\bf N. BODIES} Is the total number of particles of a simulation.}  
{\item {\bf MASS BODY} Is the mass of each particle, the unit of this quantity is given 
in eq. \ref{eq8}.}   
\end{itemize} 
Other parameters that can be changed during the run, are listed in the {\it dyn\_pars} file (see par. 5.5). 
\begin{itemize} 
{\item {\bf LIV GROU} and {\bf BODY GROU} are described in the par. 3.1.}  
{\item {\bf BOX SIZE} Is the size of the box region where the LSS simulation is executed.}  
{\item {\bf X - Y - Z MIN VER} represents the cartesian coordinates of the lowest vertex of the box.  
All the particles of the simulation must be included in the box.}  
\end{itemize} 
The remaining parameters of the {\it stat\_pars} and {\it dyn\_pars} files do not regard the physics 
of a simulation, but only the I/O filenames and the job duration, and are already described in the  
par. 5.4 and 5.5. 

\section{Running FLY}
The following procedure is recommended to perform a simulation using several \fly runs:
\begin{itemize}
\item Create the input file at the time-step cycle equal to 0 (i.e. /tmp/FLY/in\_cond\_0). We suggest to
run the FLY\_sort utility that will generate /tmp/FLY/in\_cond\_sort\_0, the new sorted input file, 
and replace it as the new input file (i.e. /tmp/FLY/posvel\_0).
\item Submit the  \fly job using the FLY\_job command file. We suggest to use the {\it ckstop} file 
that is automatically created by \fly when the executable stops the programmed time-step cycles, but the 
simulation has not yet reached the final redshift.
\end{itemize}
The User Guide gives  more details about the use of the command files.

\section{FLY scalability} 
\begin{figure}
\centering
\includegraphics[width=6cm]{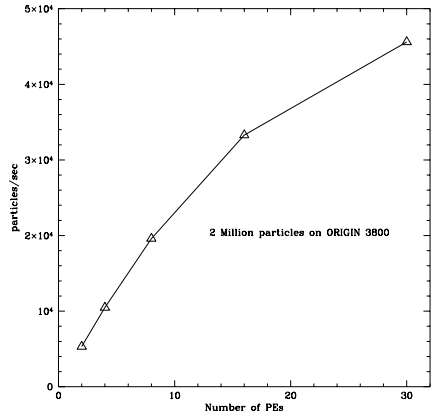}
\caption{Number of particles per sec. compute by \fly in the Sgi Origin 38000 systems,
increasing the number of PEs.}
\label{fg:sc1}
\end{figure}

\begin{figure}
\centering
\includegraphics[width=6cm]{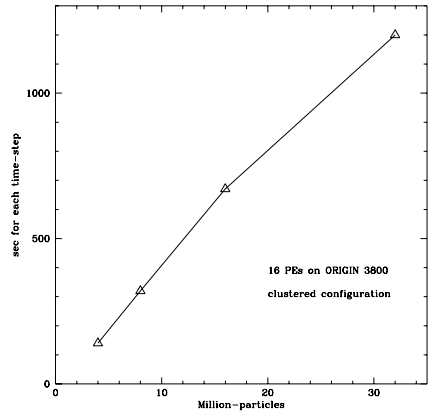}
\caption{Code scalability, increasing the number of simulated particles in a
clustered configuration and using 16 PEs.}
\label{fg:sc2}
\end{figure}

In this section we report scalability data of the \fly code using a testcase of 2097152 
particles  
in a clustered configuration using the Sgi Origin 3800 system at the Cineca. It is a 128 
processor elements (PEs) 
RISC 14000, 500 Mhz with 2 GBytes DRAM for each processor. 
The Fig. \ref{fg:sc1} shows the scalability, 
using the system from 2 PEs up to 30 PEs, the maximum PE number available for each 
run on this system. 
The Fig. \ref{fg:sc2} shows the scalability of the code. We fix the optimal number of
processor to 16 and we increase the number of particles in the simulation, from 1 million
up to 32 million-particles. All the runs were carried out using $\theta=0.8$, the quadrupole expansion,  
and a grouping level not lower than 6. More details on the \fly scalability are reported in  \cite{bec2001}. 

\section{Testcase} 
In order to check that \fly is working correctly after it has been installed, the user can use the testcase reported in the \fly distribution (Testcase direcory). Change the subdirectory of the system 
platform where you are running (i.e. ./FLY\_2.1/Testcase/SP) and create the \fly executable in this directory to run 
a test simulation with 1024 particles and 10 time-step cycles. Use the  supplied files: fly\_fnames, stat\_pars,  dyn\_pars, ew\_grid, 
ew\_tab, ql.dat, out32.tab, the input ascii file 1024\_asc\_10 and submit the job. The execution will produce the 1024\_asc\_20  
file that must be equal to {\it output\_gr} file included in the \fly distribution.
The testcase also gives another testcase with 2097152 particles: use the fly\_fnames\_2M (copy it as fly\_fnames) and create the \fly 
executable   to run it with this number of particles. In this case \fly must run without errors and the tree
properties, the number of  cells formed level-by-level, and the root cell properties (written by the \fly run in
the standard output) must be equal to the {\it output\_tree\_2M} included in the Testcase directory.

\section{FLY graphical interface}
All the input parameter files can be created using the on-line assistant that does not need  a graphical 
environment. Moreover if the user have installed the wish Tcl/Tk in the system where he
want to run \fly, he can use the graphical interface in ./FLY\_2.1/src/tcl. 
He must run {\em wish fly\_2.1.tcl} command to start
the graphical interface that help the user to create all the parameter files, 
excluding the initial condition file.\\
The first time that this interface is started, it asks for the platform, this data and other default 
data will be saved in the fly\_tcl.ini file.  
The main window sets the working directory, 
the executable directory and will create the not existing directories. 
The  figures 1 and 2 show the main window and the window to set the 
stat\_pars file. If this file exists, it is loaded with 
the values of the existing  file. Other files are created using 
similar windows.\\   
\begin{figure}
\centering
\includegraphics[width=6cm]{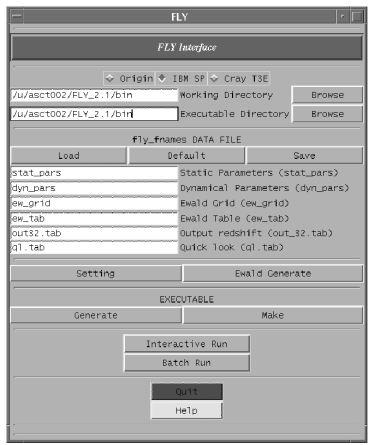}
\caption{Main FLY TCL window interface to set the fly\_fnames file}
\label{FigVibStab}
\end{figure}
   
\begin{figure}
\centering
\includegraphics[width=6cm]{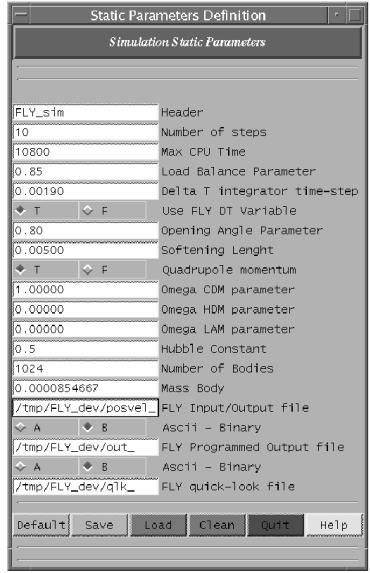}
\caption{FLY TCL window interface to set the static parameter file}
\label{FigVibStab}
\end{figure}
In the main window the user can click on the {\it Generate} button and insert  data to
create the fly\_h.F module, the makefiles  in the directory {\it src},
and  the script file that can be submitted to the system queue.
The {\it Make} button executes the makefile and creates the executable program and the FLY\_sort utility.\\
In the main window it's also possible to click on the buttons {\it Interactive Run} and {\it Batch Run}
to submit the \fly execution.

\section{FLY tools}
\fly can produce very large output files, whose analysis could be
long and cumbersome even on large computing systems. We have then included two
utilities which help the user to make some simple analysis using widespread visualization
packages, like SuperMongo and IDL. They are contained in the subdirectory {\em tools},
and are essentially a C subroutine, {\em slide\_3\_2d.c}, an IDL procedure ({\bf slice.pro}) and a SuperMongo 
macro: {\em pl\_qlk}.\\
The program {\em slide\_3\_2d.c} is an interactive utility which reads a {\em quick-look} 
input and outputs a file containing densities computed on a grid which is a 2-D projection of a 
slice of the system along the chosen line-of-sight. The latter is specified as an input parameter. 
The output is a binary file ({\bf slice.d}) containing the densities on a grid. For instance, in order 
to get  from a quick-look file {\rm qlk\_3.0}, containing $262144$ positions, a slice along the $y-$direction, 
on a grid containing $10^{3}$ cells using only particles having $15 < y < 22.4$ the command would be:
\[
\$\, {\rm slices2d -n}\, {\rm qlk\_3.0}\, {\rm -nb}\, 262144\, {\rm -nc}\, 10\, {\rm -d}\, 2\,  {\rm -min}\, 15 
\]
\[
{\rm -max}\, 22.4 
\]
The IDL procedure {\bf slice.pro} makes isodensity contours of the file 
{\bf slice.d}, while the SMongo macro {\em pl\_qlk} plots the particles on a 
given slice.\\
Another tool is the tofly.c routine executing a conversion from big-endian to 
little-endian and viceversa.\\
 
\section{Troubleshouting}
Using the assistant and the graphical interface, the user sets parameters of the fly\_h.F module and uses
it to create the \fly executable.  But many other parameters must be adjusted by the user if some
errors occur. This section describes the main errors that can occur and the 
action that must be taken to avoid the error. All the following parameters are included in the fly\_h.F module.
After any change in the parameter value, \fly must be re-compiled.\\
\begin{itemize}

\item {\bf ch\_all error 1: Abort. It is not possible to have the minimum RAM-cache size}. {\it Action:} there is not enough available 
memory for this execution. Increase the available memory if it is possible, or decrease the nb\_loc (number of local body) or 
nc\_loc (number of local cells) parameter value.
\item {\bf find\_group error 1: overflow}. {\it Action:} increase the maxilf parameter value, the maximum length of temporary storage to 
form an Interaction List.  
\item {\bf ilist error 1: overflow}. {\it Action:} increase the maxnterm parameter value.
\item {\bf ilist error 2: overflow}. {\it Action:} increase the maxilf parameter value. If the error will persist, 
please report the error to the \fly authors.
\item {\bf ilist\_group error 1: overflow}. {\it Action:} increase the maxnterm parameter value.
\item {\bf ilist\_group error 3: overflow}. {\it Action:} increase the maxilf parameter value. If the error will persist, 
please report the error to the \fly authors. 
\item {\bf read\_params error 1: nbodies greater than nbodsmax}. {\it Action:} the number value {\it nbodies} read
from the stat\_pars file doesn't match  the {\it nbodsmax} parameter. Set nbodsmax equal 
to nbodies.
\item {\bf tree\_gen error 1: Max level reached: {\it lev} greater then
nmax\_level}. {\it Action:} The number of levels created to build the tree
structure is greater than the maximum allowable levels (equal to lmax).
Increase the value of {\it lmax} parameter.
\item {\bf tree\_gen error 2: overflow}. {\it Action:} The number of cells created to build the tree
structure, is greater than the maximum allowable internal tree cells (equal to nbodsmax).
Increase the value of {\it ncells} parameter.

A complete description of all the errors that can occur is reported in the \fly User Guide.
\end{itemize}

\section{Conclusion}
\fly is still under development, so many features will be added in the future version. 
Any question,  problem and bug can be reported to the authors sending a detailed e-mail
to fly\_admin@ct.astro.it, giving a standard test problem, the input parameter files and a description of the
system where the bug was found (i.e. operating system, platform, available memory, etc.). 
An User Guide of \fly 2.1 is available with the \fly distribution \cite{flyug}. The \fly distribution 
includes a testcase for
each supported platform, allowing the user to start a simple demo of a \fly run. 
The next public version of \fly will include the computation of the gravitational
potential in an adaptive mesh like Paramesh  \cite{1999AAS...195.4203O} that
will allow the user to interface \fly outputs with
hydrodynamic codes that use adaptive mesh.
Moreover a FAQ will be prepared and will be accessible from the \fly site http://www.ct.astro.it/fly.\\



\end{document}